\documentclass[aps,prl,twocolumn,showpacs]{revtex4}
\newcommand{\bsim}{\mbox{\raisebox{-0.1cm}{$\;
\stackrel{\textstyle>}{\sim}\;$}}}
\newcommand{\lsim}{\mbox{\raisebox{-0.1cm}{$\;
\stackrel{\textstyle<}{\sim}\;$}}}
\usepackage{psfig}

\begin{document}

\title{Polaronic and nonadiabatic phase diagram from anomalous
isotope effects}

\author{P. Paci$^1$, M. Capone$^2$, E. Cappelluti$^2$,
S. Ciuchi$^3$, C. Grimaldi$^4$ and
L. Pietronero$^1$}

\affiliation{$^1$Dipart. di Fisica, Universit\`{a} di Roma ``La Sapienza",
and INFM UdR RM1, 00185 Roma, Italy}

\affiliation{$^2$``Enrico Fermi'' Research Center, c/o Compendio del Viminale,
and INFM UdR RM1, 00184 Roma, Italy}

\affiliation{$^3$Dipart. di Fisica, Universit\`{a} de L'Aquila,
and INFM UdR AQ, 67010 Coppito-L'Aquila, Italy}

\affiliation{$^4$Laboratoire de Production Microtechnique, Ecole
Polytechnique F\'ed\'erale de Lausanne, CH-1015 Lausanne,
Switzerland}

\date{\today}

\begin{abstract}
Isotope effects (IEs) are powerful tool to probe directly the dependence
of many physical properties on the lattice dynamics.  
In this paper we invenstigate the onset of anomalous IEs
in the spinless Holstein model by employing
the dynamical mean field
theory. We show that the isotope
coefficients of the electron effective mass and of the dressed
phonon frequency are sizeable also far away from the strong
coupling polaronic crossover and mark the importance of
nonadiabatic lattice fluctuations in the weak to moderate
coupling region. We characterize the polaronic regime by 
the appearence of huge IEs. 
We draw a nonadiabatic phase diagram in which 
we identify a novel crossover, not related to 
polaronic features, where the IEs
attain their largest anomalies.
\end{abstract}
\pacs{71.38.-k,63.20.Kr,71.38.Cn,71.38.Ht}
\maketitle

The relevant role of electron-phonon (el-ph) interactions on the
properties of complex materials as the high-$T_c$ superconductors
has been recently revived by a number of experiments. In
particular, the finite, and yet unexplained, anomalous IEs
on the penetration depth\cite{keller}, on the 
pseudogap temperature\cite{rubiotemprano}, and
on the angle-resolved photoemission spectra\cite{lanzara}
in high-$T_c$
superconductors open a new challenge in the understanding the
role of the el-ph interaction in these materials. Sizable isotope
effects on magnetic and charge-ordering critical temperatures in
manganites point out the relevance of the el-ph
coupling also in these materials\cite{manganiti}.

Despite that the el-ph problem has been thoroughly studied and
that many of its manifestations are now well understood, only few
partial studies have been devoted to IEs and to their
significance in relation to the underlying nature of the el-ph
interaction \cite{gcp,millis}. Yet, the prediction and observation
of IEs on different physical properties represent a
powerful tool to assess the role of the el-ph interaction in many
materials. For instance, the finite isotope shift on the
penetration depth \cite{keller}, and hence indirectly on the
effective electron mass $m^*$, is of particular interest since it
contrasts the conventional Migdal-Eliashberg (ME) scenario which
predicts strictly zero IE on this quantity. 
In this perspective the understanding of finite IEs on
$m^*$ cannot rely on the ME framework and more general approaches
are then required.

A modern tool of investigation which overcomes the limitations of 
ME theory is the so-called dynamical mean field theory (DMFT), 
a nonperturbative method which neglects spatial correlations in
order to fully account for local quantum dynamics, and becomes
exact in the infinite coordination limit\cite{revdmft}.
In the case of el-ph interactions, this approach allows us to study
with equal accuracy all coupling regimes, and to fully include
phonon quantum fluctuations which are only partially taken into account
in ME approach
according to Migdal's theorem.
DMFT has been successfully employed in the study of multiphonon effects,
polaron instabilities, metal-insulator
transitions (MIT) as well as quasiparticle regimes in the Holstein
electron-phonon system\cite{DePolarone,bulla,Max1,millis2}.

Considering the Holstein model as a paradigmatic lattice model for
el-ph interaction, we discuss the anomalies of the IEs
on electronic and phononic properties arising in
purely el-ph systems.
We employ DMFT to span the whole phase diagram determined by the
dimensionless el-ph coupling $\lambda$ and by $\gamma = \omega_0/t$,
the ``adiabatic'' ratio between the phonon energy $\omega_0$ and the 
electron hopping rate $t$. Increasing $\lambda$ gradually induces
a crossover from free carriers to polarons with reduced mobility, 
while $\gamma$ measures the relevance of quantum lattice fluctuations.
Particular attention will be paid on the dependence of
the IEs on the quantum lattice fluctuations triggered
by the finite adiabatic parameter $\gamma \neq 0$.

We show that a sizable negative isotope coefficient on the electronic mass
$m^*$ characterizes the whole parameter range, 
while a divergence of $\alpha_{m^*}$ is recovered only
approaching the MIT polaron transition at $\lambda_c$ and
$\gamma=0$.
Similar features are found for 
the renormalized phonon frequency $\Omega_0$ which displays an
isotope coefficient $\alpha_{\Omega_0}$ significantly different
from the ME limit $\alpha_{\Omega_0}=1/2$. Based on the
dependence of the isotope coefficients on the adiabatic ratio
$\gamma$, we draw a phase diagram wherein 
we identify, beside the strong-coupling polaronic regime\cite{Max1},: 
$i$) a nonadiabatic perturbative regime
where IEs increase with $\gamma$, and $ii$) a complex
nonadiabatic regime where the anomalies of the IEs
decrease with $\gamma$
and approach the Lang-Firsov predictions
in the $\omega_0/t \rightarrow \infty$ limit.

In this paper we are interested in the continuous
evolution of the
el-ph properties from the quasiparticle to the polaronic regime.
It is well known that the Holstein model undergoes various
instabilities leading to superconductivity, charge-density-wave ordering
and bipolaron formation\cite{freericks}.
In order to focus on the metallic properties and to clarify the
origin of anomalous IEs in this regime, we consider here
a half-filled spinless Holstein model, which enforces the metallic character
in the whole $\lambda$-$\gamma$ space (for $\gamma \neq 0$) \cite{Max1}.
Our Hamiltonian reads:
\begin{equation}
\label{ham}
H=-t\sum_{\langle i, j \rangle}c^\dagger_i c_j +
g\sum_i n_i(a_i+a^\dagger_i)+\omega_0\sum_i a^\dagger_i a_i,
\end{equation}
where
$c^\dagger_i$ ($c_i$), $a^\dagger_i$ ($a_i$) are creation
(annihilation) operators for electrons and phonons on site $i$, respectively,
$n_i=c^\dagger_i c_i$ is the electron density and $g$ is the el-ph
matrix element.

The physical properties of (\ref{ham})
are governed by two microscopic parameters: the el-ph coupling
$\lambda=2g^2/\omega_0 t$ and the adiabatic ratio $\gamma=\omega_0/t$.
The standard Landau Fermi-liquid (FL) picture is sustained by the ME theory
in the adiabatic limit $\gamma=0$ for $\lambda < \lambda_c$, while for
$\lambda>\lambda_c$ the FL regime is destroyed due to the polaron localization.
For $\gamma>0$, this sharp transition becomes a smooth crossover
between well and poorly defined quasiparticle properties.
This rich phenomenology is reflected in the appearances of
anomalous IEs on various quantities. 
Let us consider for example the isotope coefficient on the effective electron 
mass $m^*$: $\alpha_{m^*}=-d\ln(m^*)/d\ln(M)$, where $M$ is the ionic mass.
Since $g\propto 1/\sqrt{M\omega_0}$ and $\omega_0\propto
1/\sqrt{M}$, the el-ph coupling $\lambda$ is independent of $M$
and $\alpha_{m^*}$ can be rewritten as:
\begin{equation}
\label{iso}
\alpha_{m^*}=\frac{1}{2}\frac{d\ln (m^*/m)}{d\ln (\gamma)}.
\end{equation}
According to the FL description, $m^*$ can be expressed
in terms of a mass-enhancement factor $f_{m^*}$:
\begin{equation}
\label{LF1}
m^*/m=1+f_{m^*}(\lambda,\gamma).
\end{equation}
In the adiabatic regime $\gamma\ll 1$, $f_{m^*}$ can be expanded
in powers of $\gamma$, 
$m^*/m\simeq 1+f_{m^*}(\lambda,0)+\gamma f_{m^*}^1(\lambda)$, and
hence
\begin{equation}
\label{LF3}
\alpha_{m^*}=\frac{\gamma}{2}\frac{m}{m^*}f_{m^*}^1(\lambda).
\end{equation}
The isotope coefficient thus increases with
$\gamma$ and it correctly reproduces the ME result $\alpha_{m^*}=0$ when
$\gamma\rightarrow 0$. Such an increase is indeed found by
calculations based on a perturbative expansion in $\gamma$ \cite{gcp,millis},
whose validity is however limited only to weak  $\lambda$
values. In the opposite antiadiabatic limit $\gamma\rightarrow
\infty$ the Holstein-Lang-Firsov approximation gives
$m^*/m\simeq \exp(\lambda/2\gamma)$, leading to 
\begin{equation}
\label{HLF}
\alpha_{m^*}=-\frac{\lambda}{4\gamma},
\end{equation}
which {\it decreases} as $\gamma$ gets higher, as opposed to
to the previous case of Eq.(\ref{LF3}).
The two limiting cases for $\gamma \ll 1 $ and $\gamma \gg 1$ suggest
that for fixed $\lambda < \lambda_c$ the strongest isotope shifts lie in the
intermediate nonadiabatic region $\gamma \lsim 1$, which can be investigated
only by non-perturbative tools as the DMFT approach.

In this work we consider (\ref{ham}) on an infinite coordination
Bethe lattice and use exact diagonalization (ED) to solve the impurity problem
that DMFT associates to the lattice model\cite{revdmft,caffarel}.
As customary, the Anderson model is truncated by considering $N_s$
impurity levels, and a cutoff on the phonon number is imposed on
the infinite phonon Hilbert space.
The DMFT self-consistency is implemented in the Matsubara frequencies 
$\omega_n=(2n+1)\pi \tilde{T}$ where $\tilde{T}$ is a fictitious 
temperature.
The evaluation of IEs is a particularly difficult task since it 
requires a high accuracy on both the electron and phonon properties and of 
their dependence
on $\gamma$. In particular, a number of phonon states up to $100$ and $T/t$ as
small as $1/1600$ were needed to ensure reliable and robust results.
The number of impurity levels has been fixed at $N_s=9$, having checked that
no significant change occurred for larger $N_s$.
We compute the  electron self-energy $\Sigma(\omega_n)$
and the phonon Green's function $D(\omega_m)$
which yield the effective electron mass 
$m^*/m = 1-\Sigma(\omega_{n=0})/\pi T$
and the renormalized phonon frequency $\Omega_0$
as $(\Omega_0/\omega_0)^2 = - 2 D^{-1}(\omega_{m=0})/\omega_0$.
The corresponding isotope coefficients are obtained by means of a finite
shift $\Delta \gamma/\gamma = 0.15$.

\begin{figure}[t]
\centerline{\psfig{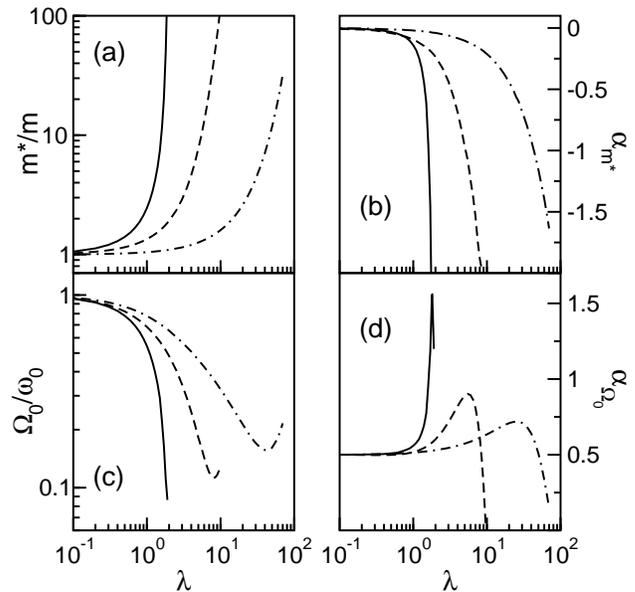}}
\caption{Effective electron mass $m^*/m$ (a) and renormalized
phonon frequency $\Omega_0/\omega_0$ (c) as function
of $\lambda$ for $\gamma=0.1$ (solid line),
$\gamma=1$ (dashed line) and $\gamma=10$ (dot-dashed line).
The corresponding isotope coefficients are shown in
panels b,d.}
\label{f-vs-lambda}
\end{figure}

In Fig. \ref{f-vs-lambda} we show $m^*/m$ and $\Omega_0/\omega_0$
and their corresponding isotope coefficients as a
function of the el-ph coupling $\lambda$ for
$\gamma=0.1, 1, 10$ which are representative respectively
of the quasi-adiabatic, nonadiabatic and antiadiabatic regimes.
The polaron crossover is reflected in a strong enhancement of $m^*/m$ as 
$\lambda$ increases. The crossover occurs at $\lambda=1.18$ for $\gamma=0$, 
and moves to larger coupling as $\gamma$ increases, until the antiadiabatic
regime, in which the crossover roughly occurs when the average number of
phonons $\alpha^2 = \lambda /2\gamma\bsim 1$\cite{Max1,DePolarone,
polaron}.
As a consequence,  at fixed
$\lambda$ the effective mass becomes smaller as $\gamma$
increases, implying a negative isotope coefficient $\alpha_{m^*}$
as reported in Fig.\ref{f-vs-lambda}(b). The polaron regime
is thus identified by the huge negative values of $\alpha_{m^*}$. Just
as in the case of $m^*/m$, increasing $\gamma$ makes the variation of 
$\alpha_{m^*}$ smoother and shifts it to larger $\lambda$.

A similar behavior is found for the
renormalized phonon frequency $\Omega_0$, as shown in
Fig.\ref{f-vs-lambda}(c), where the polaron instability is
reflected in a sharp phonon softening in the quasi-adiabatic case
($\gamma=0.1$) close to polaron crossover. Once again, the softening is weaker
as $\gamma$ gets larger. For fixed $\lambda$ this leads
to an anomalous isotope coefficient $\alpha_{\Omega_0} > 1/2$, as
reported in Fig.\ref{f-vs-lambda}(d).

Some words are worth to be spent about
the apparent phonon hardening accompanied by
the corresponding decreasing of $\alpha_{\Omega_0}$
as reported in Figs.\ref{f-vs-lambda}(c,d) for large $\lambda$.
This anomalous phonon feature
is just a consequence of our ``static'' definition of renormalized
phonon frequency $(\Omega_0/\omega_0)^2 = - 2 D^{-1}(\omega_{m=0})/\omega_0$.
This definition corresponds
to describe the full phonon spectrum as a single $\delta$-function at 
frequency $\Omega_0^2 = -1/(\pi\omega_0) \int_0^\infty d\omega
\mbox{Im}D(\omega+i0^+)/\omega$. Such description
becomes less representative as the spectral function
acquires a complex structure.

\begin{figure}[t]
\centerline{\psfig{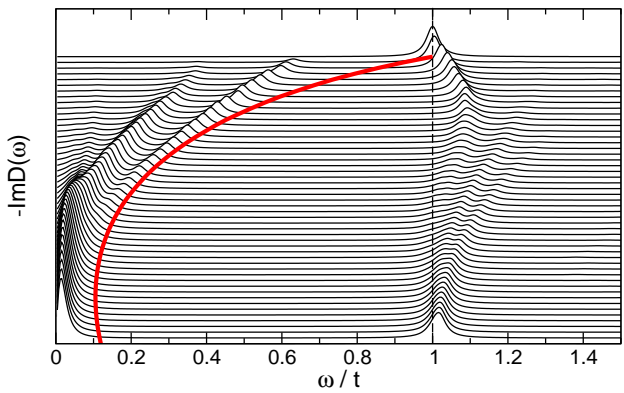}}
\caption{(color online)
Phonon spectral function (solid lines)
for different values of the el-ph coupling $\lambda$ from 0 to 10 with steps
of 0.2. The thick red line represent the renormalized phonon
frequency $\Omega_0$ and the thin dashed line $\omega_0$.}
\label{spettri}
\end{figure}

In Fig.\ref{spettri} we report the evolution of the phonon
spectral function by increasing the el-ph coupling $\lambda$
for $\gamma=1$.
The real frequency phonon propagator is directly computed
in the ED scheme. Note that only gross features can be
extracted because of the
discreteness of the impurity model. The average phonon softening
(thick red line) at small $\lambda$ stems from a transfer of
spectral weight from $\omega \simeq \omega_0$ to a low energy
peak which exponentially approaches $\omega=0$. An opposite
behavior occurs for strong el-ph coupling where the lattice
potential is a double-well with energy barrier $\Delta E \gg
\omega_0$. This gives rise to a second peak at frequency
$\omega_0$ yielding a hardening of the averaged phonon frequency
$\Omega_0$. Similar considerations lead to the decrease at large
$\lambda$ of the IE $\alpha_{\Omega_0}$.

The study of the $\lambda$ dependence of $m^*/m$, $\Omega_0$ and
their respective IEs highlights in the clearest way
the appearance of giant IEs close to the polaronic
crossover. However, interesting anomalies and sizeable deviations 
of IEs from ME predictions  
appear far from polaronic regime in a region where the system preserves good 
metallic properties. These anomalous features can
thus be connected to the onset of nonadiabatic effects in the
el-ph properties in the spirit of a generalization of the FL
liquid picture as in Eq.(\ref{LF1})\cite{gps,gcp,cgps}. 
To this aim we analyze in
more details the dependence of the IEs on the quantum
lattice fluctuations triggered by the finite adiabatic parameter
$\gamma \neq 0$.
\begin{figure}[t]
\centerline{\psfig{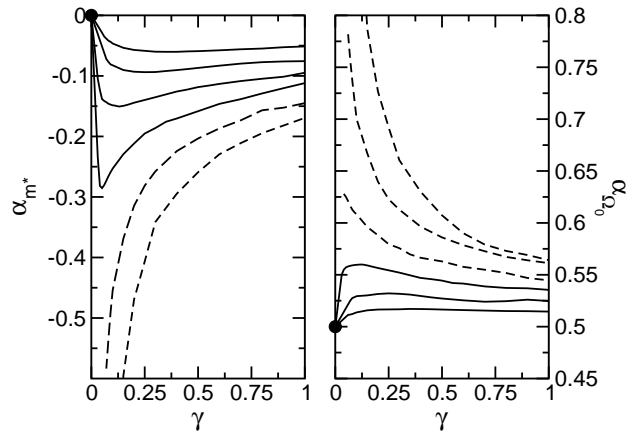}}
\caption{Dependence of $\alpha_{m^*}$ and $\alpha_{\Omega_0}$ on
the nonadiabatic parameter $\gamma$. Curves are plotted for the
el-ph coupling $\lambda=0.6, 0.8, 1.0, 1.2, 1.4, 1.6$ from top to
bottom (bottom to top) in left (right) panel. Filled black
circles mark the adiabatic ME limit.} \label{iso-w0}
\end{figure}
In Fig. \ref{iso-w0} we plot the $\gamma$-dependence of
$\alpha_{m^*}$ and $\alpha_{\Omega_0}$ for different el-ph
coupling ranging from weak ($\lambda=0.6$) to intermediate-strong
coupling ($\lambda=1.6$). Two qualitatively different behaviors
marked by the solid and dashed lines are identified. Dashed lines
represent strong coupling $\lambda$ values for which the system
undergoes a polaronic MIT for  $\gamma \to 0$
signaled by the divergence of the isotope coefficients.
Solid lines are representative of
weak-to-moderate $\lambda$ values for which the system maintains
its FL metallic character with isotope coefficients recovering
their standard ME values $\alpha_{m^*}=0$ and
$\alpha_{\Omega_0}=1/2$ as $\gamma\rightarrow 0$. Note that for
$\lambda=1.2$ we find that $\alpha_{m^*}\rightarrow 0$ and
$\alpha_{\Omega_0}\rightarrow \infty$ for $\gamma\rightarrow 0$,
in agreement with Ref. \cite{millis2} which predicts two
different critical values $\lambda_c^{\rm ph}=1.18$ and
$\lambda_c^{\rm el}=1.328$ where a double-well lattice potential
and an electronic self-trapping respectively occur.

The different physics underlying the regimes sketched by dashed and solid lines
is reflected in the overall $\gamma$ dependence of the IEs. In the
large coupling regime $\lambda > \lambda_c$ the isotope
coefficients diverge for $\gamma \rightarrow 0$ and approach
monotonically zero in the antiadiabatic limit. For  $\lambda <
\lambda_c$ on the other hand we find a more complex behavior with
an initial increase of the anomalies of $\alpha_{m^*}$ and
$\alpha_{\Omega_0}$ as function of $\gamma$ up to a certain value
$\gamma^*$, above which the anomalies 
decrease and disappear in the antiadiabatic $\gamma \rightarrow
\infty$ limit in agreement with the Lang-Firsov predictions.
In the nonadiabatic crossover region
($\lambda<\lambda_c$ and $\gamma\sim\gamma^*$) significant anomalies
of the IEs $\alpha_{m^*}\sim -0.3$ and
$\alpha_{\Omega_0}\sim 0.55$ are observed even in the metallic
regime far from the polaronic instability. It is worth to stress
that the values of $\gamma^*$ strongly depends on the el-ph
coupling $\lambda$ and that in principle a nonadiabatic crossover
parameter can be defined for both the electronic and phononic
properties.

As discussed above, the curve $\gamma^*$ separates a region in which a
perturbative theory based on the adiabatic FL picture can be safely employed
($\gamma < \gamma^*$) from a region where the Lang-Firsov approach is a more
appropriate starting point for a $1/\gamma$ expansion ($\gamma > \gamma^*$).
However around $\lambda_c$ at $\gamma\simeq 0$ both ME and Lang-Firsov
approaches are expected to fails due to strong entanglement of electron and
phonon degrees of freedom \cite{DePolarone}.

\begin{figure}[t]
\centerline{\psfig{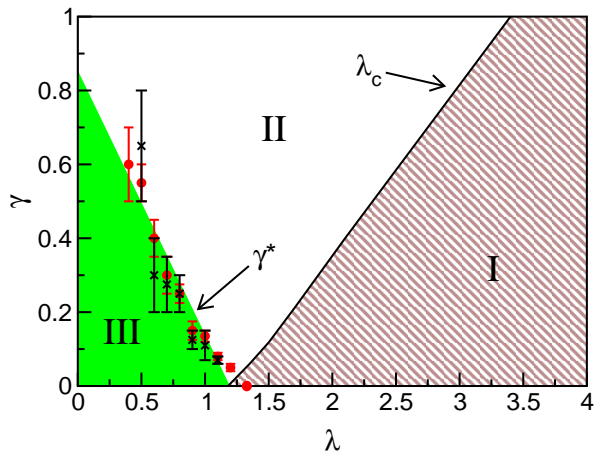}}
\caption{(color online) Nonadiabatic phase diagram defined by
anomalous IEs. Filled red circles and black crosses
represent respectively the minima and the maxima in the $\gamma$
behavior of $\alpha_{m^*}$, $\alpha_{\Omega_0}$. Error bars are
not related to the accuracy of the isotope coefficients but more
to the flatness of $\alpha_{m^*}$, $\alpha_{\Omega_0}$ as
function of $\gamma$. The curve $\lambda_c$ marks the polaron
crossover as defined in the text.} \label{diagram}
\end{figure}

The complex phenomenology of the IEs in the whole
$\lambda$-$\gamma$ phase diagram is summarized in Fig.
\ref{diagram}, where the nonadiabatic crossover $\gamma^*$ is
defined respectively by the minima (filled circles) and maxima
(crosses) of the $\alpha_{m^*}$, $\alpha_{\Omega_0}$ as functions
of $\gamma$.
We note that, apart from a very small region close to the
adiabatic limit, $\gamma^*$ lies on an universal curve for both
the electron and phonon properties reflecting the fact that
electron and lattice degrees of freedom are strictly mixed in the
nonadiabatic regime. This is indeed not true for $\gamma
\rightarrow 0$ where $\gamma^*$ disappear at the two different
values $\lambda_c^{\rm ph}$, $\lambda_c^{\rm el}$ above discussed.
It is also worth to stress that the nonadiabatic crossover,
pointed out by the $\gamma$ dependence of the anomalous isotope
effects, is not related to the onset of polaronic effects,
as shown in Fig. \ref{diagram} where
the polaron crossover at finite $\gamma$ is here pinpointed 
by the appearance of a bimodal
structure in the lattice probability distribution function. For
$\lambda > \lambda_c$ the polaron crossover is reflected in a
strong degradation of the metallic properties with an almost
vanishing coherent spectral weight. The behaviors of $\lambda_c$
and $\gamma^*$ define three regimes which can be roughly
described as: I) a polaron region (brown dashed area in Fig.
\ref{diagram}) where electrons are almost trapped leading to
significant lattice distortions. This effect is strongly
removed by the lattice quantum fluctuation triggered by a finite
$\omega_0$ leading to giant IEs; II) a highly
nonadiabatic region (white area) where the system is
qualitatively described in terms of itinerant quasi-particle
carrying along its hugely fluctuating phonon cloud. In this
regime quantum lattice fluctuations relax the mixing between
lattice and electronic degrees of freedom leading to a reduction
of the anomalous IEs; III) a weakly nonadiabatic
region (grey filled area) where anomalous IEs are
tuned by the opening of nonadiabatic channels in the el-ph
interaction. In this region DMFT qualitatively confirms the
results of the nonadiabatic theory described by vertex diagrams
in a perturbative approach\cite{gps,gcp}.

In conclusion we have defined a phase diagram of the spinless Holstein model
based on the anomalous phenomenology of the IEs on
both electronic and phonon properties. In the metallic regime we
identified a new nonadiabatic crossover $\gamma^*$ (not related to polaronic
instability) between different weakly and highly nonadiabatic FL quasi-particle
picture. The largest anomalies in the isotope coefficients are found
in this intermediate crossover region. Experimental investigations
of anomalous IEs on different physical properties
represent thus a powerful tool to probe the complex nature of
the el-ph interaction in real materials. Theoretical studies based on
Density Functional Theory predict $A_3$C$_{60}$ fullerides,
cuprates and MgB$_2$ to be roughly in the nonadiabatic crossover
where the anomalies in the IEs are the most sizable.
A rigorous generalization of our results in the presence of
electronic correlation and away from the half-filling case is
of course needed for quantitative analysis in these complex materials.
Preliminary results show that $\alpha_{m^*}$ can be significantly suppressed
in low filled systems even in highly nonadiabatic regime, 
in agreement with recent experimental results
in MgB$_2$\cite{mgb2}.

We are grateful to C. Castellani for  illuminating discussions.
We acknowledge financial support from Miur Cofin2003 and FIRB RBAU017S8R and  
INFM PRA-UMBRA.

\end{document}